\documentclass[12pt]{article}
\pdfoutput=1
\usepackage{array} 
\usepackage{amssymb}
\usepackage{graphics,graphpap}
\usepackage{graphicx}
\usepackage{color}
\usepackage{graphicx}
\usepackage{dcolumn}
\usepackage{epsfig}
\usepackage{epstopdf}
\usepackage{bbm}
\usepackage{amsmath}
\usepackage{amsfonts}
\usepackage{textcomp}
\usepackage{subcaption}
\usepackage{setspace}
\usepackage{slashed}
\usepackage{multirow}
\usepackage{footnote}
\makesavenoteenv{table}
\usepackage{url}
\usepackage{slashed}
\usepackage{mathtools}

\usepackage{tikz}
\definecolor{dunkelgruen}{rgb}{0,0.7,0}
\definecolor{dunkelblau}{rgb}{0,0,0.7}

\usepackage{xcolor}
\definecolor{red}{rgb}{1,0,0}
\definecolor{purple}{rgb}{0.5,0,0.5}
\definecolor{blue}{rgb}{0,0,1}

\newcommand{\beq}{\begin{eqnarray}}
\newcommand{\eeq}{\end{eqnarray}}

\newcommand{\bmp}{\noindent\begin{minipage}{16cm}}
\newcommand{\emp}{\end{minipage}\vskip 7mm} 

\newcommand{\ra}{\rightarrow}

\def\drawbox#1#2{\hrule height#2pt
        \hbox{\vrule width#2pt height#1pt \kern#1pt
              \vrule width#2pt}
              \hrule height#2pt}

\def\Asym#1#2{\vcenter{\vbox{\drawbox{#1}{#2}
              \kern-#2pt 
              \drawbox{#1}{#2}}}}

\def\simge{\mathrel{%
   \rlap{\raise 0.511ex \hbox{$>$}}{\lower 0.511ex \hbox{$\sim$}}}}

\def\simle{\mathrel{
   \rlap{\raise 0.511ex \hbox{$<$}}{\lower 0.511ex \hbox{$\sim$}}}}

\def\s#1{\setbox0=\hbox{$#1$}%
\rlap{\ifdim\wd0>.7em\kern.22\wd0\else\kern.1\wd0\fi /}#1}

\definecolor{purple}{rgb}{0.5,0,0.5}

\usepackage{cite}

\begin{document}

\begin{titlepage}
\title{\vspace*{-2.0cm}
\hfill {\small MPP-2016-298}\\[20mm]
\vspace*{-1.5cm}
\bf\Large
$\boldsymbol{\mu^-}$-- $\boldsymbol{e^+}$ conversion in upcoming LFV experiments \\[5mm]\ \vspace{-1cm}}

\author{
Tanja Geib$^a$\thanks{email: \tt tgeib@mpp.mpg.de}~~~,~~Alexander Merle$^a$\thanks{email: \tt amerle@mpp.mpg.de}~~,~~and~~Kai Zuber$^b$\thanks{email: \tt zuber@physik.tu-dresden.de}
\\ \\
$^a${\normalsize \it Max-Planck-Institut f\"ur Physik (Werner-Heisenberg-Institut),}\\
{\normalsize \it F\"ohringer Ring 6, 80805 M\"unchen, Germany}\\
\\
$^b${\normalsize \it Institute for Nuclear and Particle Physics,}\\
{\normalsize \it Technische Universit\"at Dresden, Germany}\\
}
\date{\today}
\maketitle
\thispagestyle{empty}

\begin{abstract}
\noindent
We present an illustrative discussion of the physics potential of $\mu^-$-- $e^+$ conversion. We point out that this process, although much less frequently studied than the related but much more popular processes of $\mu^-$-- $e^-$ conversion and neutrinoless double beta decay, in fact is a promising alternative possibility to detect both lepton flavour and number violation. However, for this goal to be reached, a combined effort of experiments and theory, both in nuclear and particle physics, is necessary to advance. The aim of this paper is to be an ``appetiser'' to trigger such an initiative.
\end{abstract}

\end{titlepage}

\section{\label{sec:intro}Introduction}

A detection of lepton number violation (LNV) would shake the fundaments of our current particle physics picture. In the Standard Model (SM), lepton number can only be violated by non-perturbative processes which do not appear at low energies~\cite{'tHooft:1976up,Klinkhamer:1984di}, while any perturbative process (i.e., any Feynman diagram) does not lead to LNV. For example, in the SM, we have no possibility to write down a diagram for neutrinoless double beta decay ($0\nu\beta\beta$), $(Z,A) \to (Z+2,A) + 2e^- $, in which a nucleus with atomic number $Z$ and mass number $A$ decays such that, although no lepton has been present in the initial state, two electrons appear in the final state. However, many new physics theories beyond the SM do feature LNV, which is why its detection would be of such great importance.

If the physics community were to bet on how to observe LNV, most experts would go with $0\nu\beta\beta$. Indeed, many experiments are currently searching for this process. Among them are GERDA~\cite{Agostini:2016iid}, EXO-200~\cite{Albert:2014awa}, KamLAND-Zen~\cite{Gando:2012zm}, and many more, and the limits on the half-life have reached an impressive level in far excess of $10^{25}$~yrs. This is of course due to tremendous progress on the experimental side within the last decade, however, in parts this success also comes from the possibility to use large amounts of isotopically enriched material -- which have the benefit of exploiting the large number of atoms contained in a macroscopic piece of matter. Yet, after all, $0\nu\beta\beta$ can \emph{only} detect LNV if it is present in the $ee$-sector, because the process is energetically only possible if electrons are involved.

But exactly that may be a problem: nobody can guarantee that LNV is in fact most prevalent in the $ee$-sector. On the contrary, looking at the literature (see, e.g., Refs.~\cite{Chen:2006vn,Cirigliano:2004tc,King:2014uha}), there are several settings known in which by far ``more'' LNV is present in, e.g., the $e\mu$ or $e\tau$ sector, while it is strongly suppressed or even completely switched off in the $ee$-channel. What can be done in such a case? Indeed, the possibilities are considerably more scarce. Currently, some limits do exist stemming from rare kaon decays like  BR$(K_L^+ \to  e^\mp \mu^\pm) <4.7\cdot 10^{-12}$\ by E871~\cite{Ambrose:1998us},  BR$(K^\pm \to  \pi^\mp \mu^\pm \mu^\pm) <8.6\cdot 10^{-11}$@90\%~C.L.\ by NA48~\cite{Massri:2016dff}, as well as exotic meson decays, e.g.\ BR$(D^+ \to K^- e^+ \mu^+) <1.9\cdot 10^{-6}$@90\%~C.L.\ by BaBar~\cite{Lees:2011hb} or BR$(B^+ \to D^- e^+ \mu^+) <1.8\cdot 10^{-6}$@90\%~C.L.\ (by BELLE~\cite{Seon:2011ni}), or from rare tau decays, e.g.\ BR$(\tau^- \to e^+ \pi^- \pi^-) <2.0\cdot 10^{-8}$@90\%~C.L.\ (by BELLE~\cite{Miyazaki:2012mx}). However, these are not even remotely close to the figures characteristic for $0\nu\beta\beta$. In the future, one could probably make the greatest progress with a new lepton collider (see, e.g., Ref.~\cite{Rodejohann:2010bv}), however, it is uncertain whether such a machine will ever be built.

We therefore make a case for the alternative LNV process (or, more specific, charged lepton number \emph{and} flavour violating -- CLNFV) of bound $\mu^-$-- $e^+$ conversion. While it had already been proposed decades ago~\cite{Pontecorvo:1967fh,Kisslinger:1971vw,Shuster:1973se}, it is only now that experiments on the similar but only lepton \emph{flavour} violating (LFV) process of coherent $\mu^-$-- $e^-$ conversion are expected to increase their sensitivity on the branching ratio by several orders of magnitude~\cite{Raidal:2008jk} -- possibly even reaching an incredible sensitivity of BR$[\mu^- + (Z,A) \to e^- + (Z,A)] = \mathcal{O}(10^{-18})$~\cite{Barlow:2011zza}. This is crucial because, for most experiments aiming to measure LFV $\mu^-$-- $e^-$ conversion, the additional measurement of the LNV $\mu^-$-- $e^+$ conversion comes practically \emph{for free} -- or with very minor modifications of the setup. Thus, with sensitivities not identical but at least similar to those on $\mu^-$-- $e^-$ conversion, we can also expect an improvement on the bounds on $\mu^-$-- $e^+$ conversion by several orders of magnitude within the coming years. This CLNFV conversion has been targeted in previous experiments~\cite{Bryman:1972rf,Abela:1980rs,Badertscher:1981ay,Burnham:1987gr,Ahmad:1988ur,Kaulard:1998rb,Bertl:2006up}, however, nowadays most of this expertise seems to be ``lost'', and it is worth reconsidering $\mu^-$-- $e^+$ conversion in the light of the newest technology.

In the following, we will illustrate that $\mu^-$-- $e^+$ conversion can possibly be used to gain fundamental physics insights. We will clearly single out the three directions in which advances are necessary to ensure this progress: the more detailed investigation of particle physics models in what regards LNV in the $e\mu$ sector (to understand the possible gain), more involved experimental sensitivity studies (to determine the physics potential of upcoming experimental setups), and the up-to-now missing computation of the nuclear matrix elements (NMEs) for the process (to tighten the resulting limits on promising theories). Note that, in the first point, we anticipate some of the results of a detailed on-going study aiming to determine the contributions of a set of certain particle physics models to the short-range operators transmitting $\mu^-$-- $e^+$ conversion~\cite{Geib:2016daa}.

\section{\label{sec:formal}Formalism}

\begin{figure}[t]
\centering
\begin{tabular}{lcr}
\begin{minipage}{7cm}
\includegraphics[width=7cm]{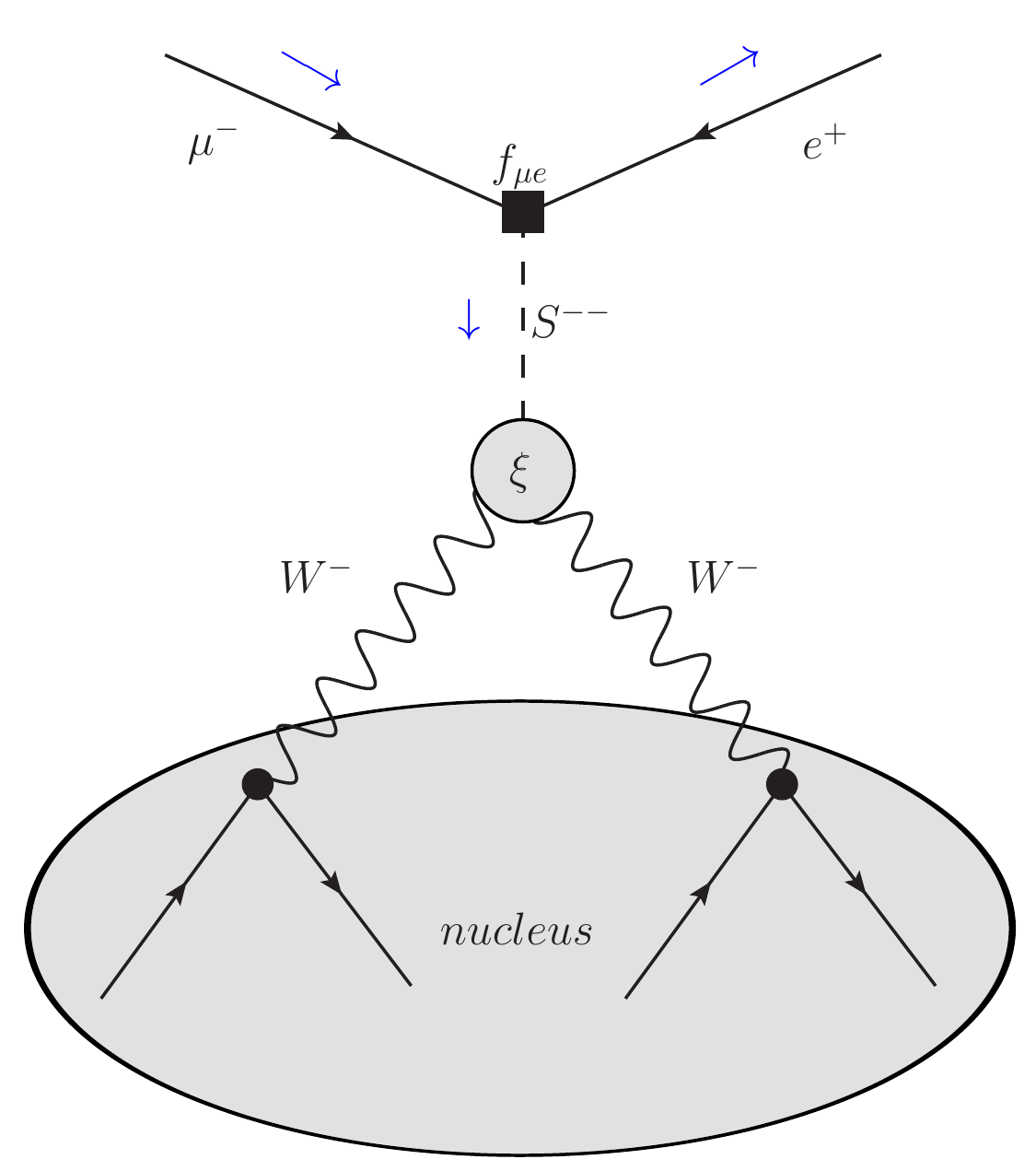}
\end{minipage}
&
\begin{minipage}{3cm}
{\Huge $\Leftrightarrow$}\vspace{1.0cm}
\end{minipage}
&
\begin{minipage}{8cm}
\hspace{-2cm}
\includegraphics[width=8cm]{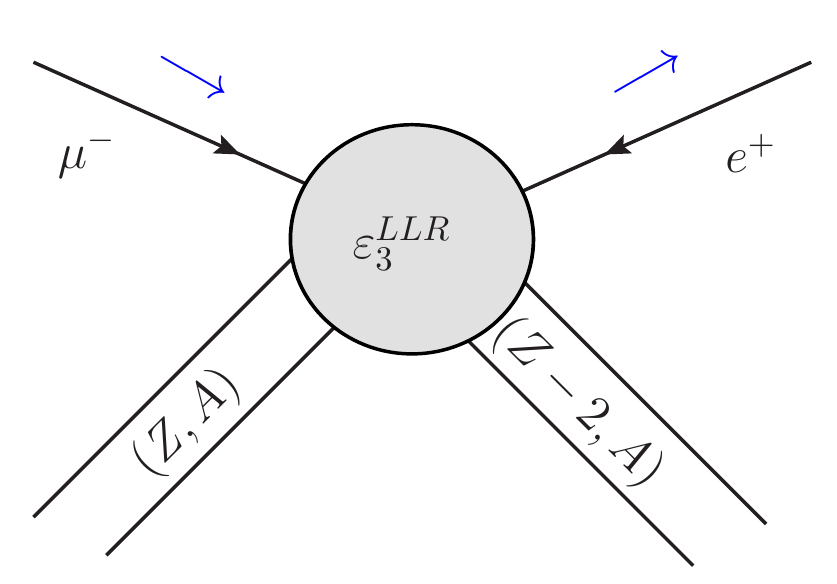}
\end{minipage}
\end{tabular}
\caption{\label{fig:Example_2}Model with a doubly charged scalar $S^{--}$ mapped onto a short-range operator, cf.\ Eq.~\eqref{eq:short_ex_2}. Blue arrows indicate the direction of momentum.}
\end{figure}

In order to consider the short-range contributions to the $\mu^-$-- $e^+$ conversion within a general framework, we turn to an effective field theory treatment analogous to the one used for neutrinoless double beta decay ($0\nu\beta\beta$)~\cite{Pas:2000vn}, which covers all \emph{short-range} contributions. Hence, the bound muon and the positron interact with the nucleons via point-like vertices. We restrict ourselves to the short-range operators of lowest dimension, $d=9$. Taking into account Lorentz invariance, the most general short-range Lagrangian is~\cite{Pas:2000vn}:\footnote{The corresponding EFT parametrisation for the long-range part, which is needed if e.g.\ light Majorana neutrinos realise the conversion, will not be included in the following discussion. The long-range contributions can be parametrised in analogy to $0\nu \beta \beta$, though, see \cite{Pas:1999fc} for a thorough discussion.}
\begin{eqnarray}
&&\mathcal{L}_\text{short-range}^{e\mu} = \frac{G_F^2}{2 m_p} \sum_{x,y,z = L,R}\big[ \epsilon_1^{xyz} J_x J_y j_z + \epsilon_2^{xyz} J_x^{\nu \rho} J_{y,\nu \rho} j_z + \epsilon_3^{xyz} J_x^\nu J_{y,\nu} j_z + \epsilon_4^{xyz} J_x^\nu J_{y,\nu \rho} j_z^\rho\nonumber\\
 && +  \epsilon_5^{xyz} J_x^\nu J_y j_{z,\nu} + \epsilon_6^{xyz} J_x^\nu J_y^\rho j_{z,\nu\rho} + \epsilon_7^{xyz} J_x J_y^{\nu \rho} j_{z,\nu\rho} + \epsilon_8^{xyz} J_{x,\nu \alpha} J_y^{\rho \alpha} j_{z,\rho}^\nu \big]\,,
 \label{eq:short-range}
\end{eqnarray}
where $G_F=\sqrt{2}g^2/(8M^2_W)$ is the Fermi constant and $m_p$ is the proton mass. The hadronic currents are defined similarly as in Ref.~\cite{Bergstrom:2011dt}:
\begin{equation}
 J_{R,L} = \overline{d} (1 \pm \gamma_5) u, \ \ J_{R,L}^\nu = \overline{d} \,\gamma^\nu(1 \pm \gamma_5) u,\ \ J_{R,L}^{\nu\rho} = \overline{d}\, \sigma^{\nu\rho} (1 \pm \gamma_5) u\,.
 \label{eq:hadronic-currents}
\end{equation}
The leptonic currents are defined analogously, however, connecting $\mu$-$e$ instead of $e$-$e$:
\begin{equation}
\begin{split}
  j_{R,L} &= \overline{e^c} (1 \pm \gamma_5) \mu = 2 \overline{(e_{R,L})^c}\, \mu_{R,L},\ \ j_{R,L}^\nu = \overline{e^c}\, \gamma^\nu (1 \pm \gamma_5) \mu = 2 \overline{(e_{L,R})^c} \,\gamma^\nu \mu_{R,L} \\ 
  {\rm and} & \ \ j_{R,L}^{\nu\rho} = \overline{e^c}\, \sigma^{\nu\rho} (1 \pm \gamma_5) \mu=2 \overline{(e_{R,L})^c}\,\sigma^{\nu\rho} \mu_{R,L}\,.
\end{split}
 \label{eq:leptonic-currents}
\end{equation}
Depending on the nature of the LNV physics, one or the other operator may be realised, and a bound on the very same operator can have different implications depending on which model generates it. Note that, while for $0\nu\beta\beta$ the operators with coefficients $\epsilon_{6,7,8}$ can be shown to vanish due to the anti-symmetry of operators connecting two electron fields~\cite{Prezeau:2003xn}, this logic does not hold anymore when different flavours are combined. However, as we will explicitly demonstrate in~\cite{Geib:2016daa}, one can show that these operators do not contribute in the limit of perfectly non-relativistic nucleons, which is generally a rather good approximation and which implies that $\epsilon_{6,7,8}$ will only contribute as higher-order corrections. Note further that, although the operators in Eq.~\eqref{eq:short-range} can appear in very different chirality structures, in most cases the experimental limit depends much more on the index $n$ of $\epsilon_n$ rather than on which chiral structure is realised~\cite{Geib:2016daa,Pas:2000vn,Bergstrom:2011dt,Gonzalez:2015ady}. In many realistic settings, however, only one or a few of the short-range operators from Eq.~\eqref{eq:short-range} are realised. For example, a doubly charged singlet scalar as introduced in Ref.~\cite{King:2014uha} would only admit the single operator:
\begin{equation}
 J_L^\mu J_{L,\mu} j_R:\ \ \ \epsilon_3^{LLR} = 4 V^2_{ud}\,m_p\,\frac{f^*_{e\mu} v^4\,\xi}{\Lambda^3 M^2_S},
 \label{eq:short_ex_2}
\end{equation}
where $\xi$ is a lepton number violating effective coupling, $v=246$~GeV is the vacuum expectation value of the SM Higgs, $f_{e\mu}$ is the lepton flavour violating coupling of the charged singlet scalar with mass $M_S$ to charged right-handed leptons, and $\Lambda$ is the ultra-violet cutoff of the model considered in~\cite{King:2014uha}. For illustrative purposes, we have depicted the mapping onto the short-range operator in Fig.~\ref{fig:Example_2}. Treating the short-range contributions via an EFT allows for the separation of the nuclear physics part from the respective particle physics model. It thereby allows for the (particle-) model-independent computation of the NMEs and, thus, a wide range of particle physics models can be attacked by a single strike. Consequently, it is essential to determine the relevant $\mu^-$-- $e^+$ conversion NMEs, such that limits from this CLNFV process can be derived. In case only a short-range operator of type $\epsilon_3^{xxz}$\footnote{Note that in case $\epsilon^{xyz}_3$ with $x\neq y$ is realised, the NME takes a slightly different form, i.e., there is a relative sign change in between the Gamow-Teller and the Fermi contributions in comparison to $x=y$. For further details, see Ref.~\cite{Geib:2016daa}.} is realised, the decay rate is given by:\footnote{Note that this decay rate differs by a factor of $\pi$ from the one obtained in~\cite{Domin:2004tk}. For more details on the derivation of the decay rate and the formalism used see~\cite{Geib:2016daa}.}
\begin{equation}
 \Gamma = \frac{1}{32\pi^2}\,G_F^4\,g^4_A\,\big|\epsilon_3^{xxz} \big|^2\,\frac{m^2_e\,m^2_\mu}{R^2}\,\big|F(Z-2,E_e)\big| \,\langle \phi_\mu \rangle^2\,\big|\mathcal{M}^{(\mu^-,e^+)}\big|^2\,,
 \label{eq:DecayRate}
\end{equation}
where $g_A=1.254$~\cite{Domin:2004tk} and $R=1.1 A^{1/3}$~fm is the nuclear radius for an atom with mass number $A$. Here, $\mathcal{M}^{(\mu^-,e^+)}$ is the NME as defined in Eq.~(49) of \cite{Domin:2004tk}, and $\langle \phi_\mu \rangle^2=\frac{\alpha^3 m^3_\mu}{\pi}\frac{Z_{\text{eff}}}{Z}$ approximates the muon average probability density~\cite{Kosmas:1993ch}. The Fermi function $F(Z-2,E_e)$ is introduced to account for the influence of the nucleus' Coulomb potential on the final state positron. From Eq.~\eqref{eq:DecayRate} it is evident that particle physics models realising some form of the short-range operator coefficient $\epsilon_3^{xxz}$ can be constrained by a non-observation of the process, as we will illustrate in Fig.~\ref{fig:COMET-Reach}. Let us briefly discuss some more examples.

In case the transition is mediated by heavy right-handed Majorana neutrinos, as discussed in Ref.~\cite{Simkovic:2000ma}, the operator realised is:
\begin{equation}
 J_L^\mu J_{L,\mu} j_L:\ \ \ \epsilon_3^{LLL} = 2 V^2_{ud}\,m_p\,\langle M^{-1}_N \rangle_{e\mu}\,,
 \label{eq:short_ex_1}
\end{equation}
where $\langle M^{-1}_N \rangle_{e\mu} \equiv \sum_k \frac{U_{ek} U_{\mu k}}{M_k}$ is the effective (inverse) mass parameter describing how the active-neutrino flavours $e$ and $\mu$ mix with the heavy right-handed neutrinos $N_k$.

Another model realising only $\epsilon_3$ is a Left-Right symmetric model supplemented by additional Higgs bosons and singlet fermions as put forward in Ref.~\cite{Pritimita:2016fgr}. The resulting effective short-range operator and coupling are:
\begin{equation}
 J_R^\mu J_{L,\mu} j_L:\ \ \ \epsilon_3^{RLL} = V_{ud}^2 \frac{\langle p \rangle^3}{m_e} \left( \frac{g_R}{g_L} \right) \sum_{j=1}^3 \frac{V_{ej}^{\nu S} V_{\mu j}^{SS}}{M_{S_j}^2} \tan \zeta_{LR}\,,
 \label{eq:short_ex_LR}
\end{equation}
where $\zeta_{LR}$ is the $W$-boson mixing angle, $g_R \sim g_L$ are the $SU(2)_{R,L}$ gauge couplings, and $M_{S_j}$ denotes the mass of the respective singlet fermion $S_j$ contained in the model; furthermore, $\langle p \rangle \simeq 100$~MeV is the average nucleon momentum scale whose definition includes the factor of $m_p$ that would otherwise appear in $\epsilon_3^{RLL}$, see~\cite{Pritimita:2016fgr} for details. Although this example does not fulfil the requirements of using Eq.~(\ref{eq:DecayRate}) in combination with the NME derived in Ref.~\cite{Domin:2004tk}, we will include it in the following discussion and estimate its NME to be of the same order of magnitude. This approach is justified by considering the NMEs of $0\nu \beta \beta$ for which this estimate holds~\cite{Pas:2000vn}. Furthermore, we only aim at ballpark estimates on the discovery potential for $\epsilon_3$ anyway because the NME values are uncertain, see Sec.~\ref{sec:process}.

Our final example is taken from the two-loop neutrino mass model of Ref.~\cite{Chen:2006vn}, where the SM is extended by an $SU(2)$ triplet and a doubly charged scalar. The doubly charged component of the triplet and the singlet scalar mix to physical mass eigenstates $P_{1,2}^{\pm \pm}$ which realise the conversion via:
\begin{equation}
 J_L^\mu J_{L,\mu} j_R:\ \ \ \epsilon_3^{LLR} = \frac{4 m_p V^2_{ud}}{\sqrt{2}} Y_{e\mu} |\sin (2\omega)| \left| \frac{1}{M_1^2} - \frac{1}{M_2^2} \right|\;,
 \label{eq:short_ex_Ng}
\end{equation}
where $\omega$ is the mixing angle of the doubly charged scalar mass eigenstates of masses $M_{1,2}$, $v_T$ is the vacuum expectation value of the triplet Higgs, and $Y_{e\mu}$ denotes the singlet Yukawa coupling to two charged right-handed leptons.

Yet another class of models that generate LNV are those based on $R$-parity violating (RPV) supersymmetry (SUSY). Within the framework of RPV-SUSY, there are several mechanisms that provide LNV which are discussed broadly in the literature, e.~g.~\cite{Faessler:1997db,Faessler:2011qw,Bergstrom:2011dt}, for the case of $0\nu\beta\beta$. While we focus on short-range operators here, there are also interesting long-range contributions that can lead to sizeable contributions (see e.g.\ Ref.~\cite{Babu:1995vh}, where the cases of sbottom/stau exchanges avoid stringent bounds on the SUSY parameter space). When contemplating RPV SUSY, there are several mechanisms that provide $\mu^-$-- $e^+$ conversion. For models with neutralino exchange being dominant~\cite{Faessler:1997db}, the contribution is similar to that from heavy right-handed neutrinos, cf.\ Eq.~\eqref{eq:short_ex_1}. Although in this case $\epsilon_3$ is realised, such that limits from experiment can be translated, we do not consider it in the following due to its smallness. A general problem with these RPV models is that, although potentially promising, they cannot be properly assessed at the moment -- which is why we could not include them in our analysis. For example, the short-range contributions discussed in~\cite{Faessler:2011qw} only realise the operators $\epsilon_1$ and/or $\epsilon_2$, and the same is true for the particularly promising long-range operators proposed in~\cite{Babu:1995vh}. Thus, even though we can evaluate the effective operator coefficients in these settings and they seem to be rather large (possibly even larger than the ones we have included in Fig.~\ref{fig:COMET-Reach}), at the moment no computation of the corresponding NMEs is available for these cases. Therefore, we are unable to give a reliable prediction on how strongly these promising contributions could be constrained in the future. This is one particular example of advances being necessary on the nuclear physics side, and it may possibly motivate nuclear physics theorists to spread out their techniques of computing NMEs to cases involving $\mu^-$-- $e^+$ conversion.

For the time being, though, when considering $\mu^-$-- $e^+$ conversion, we have to rely on the computations that exist at this stage. In fact, several authors have recognised the potential benefits of this process, so that the NMEs for the short- and long-range operators corresponding to $\epsilon^{xxz}_3$ are already available~\cite{Domin:2004tk}. We will thus start with this case, for which no striking but at least several interesting cases exist. But, in order to fully exploit the discovery potential that lies within the next generation of groundbreaking bound muon experiments like COMET~\cite{COMET}, DeeMe~\cite{Aoki:2010zz}, or Mu2e~\cite{Kutschke:2011ux}, we are in dire need of the nuclear physics community advancing on the yet unknown NMEs, and hopefully our first investigations can act as initial spark for further detailed studies.

\section{\label{sec:process}Experimental aspects of $\mu^-$-- $e^+$ conversion}

\begin{figure}[t]
  \centering
  \includegraphics[trim=0cm 0cm 0cm 8cm,width=11cm]{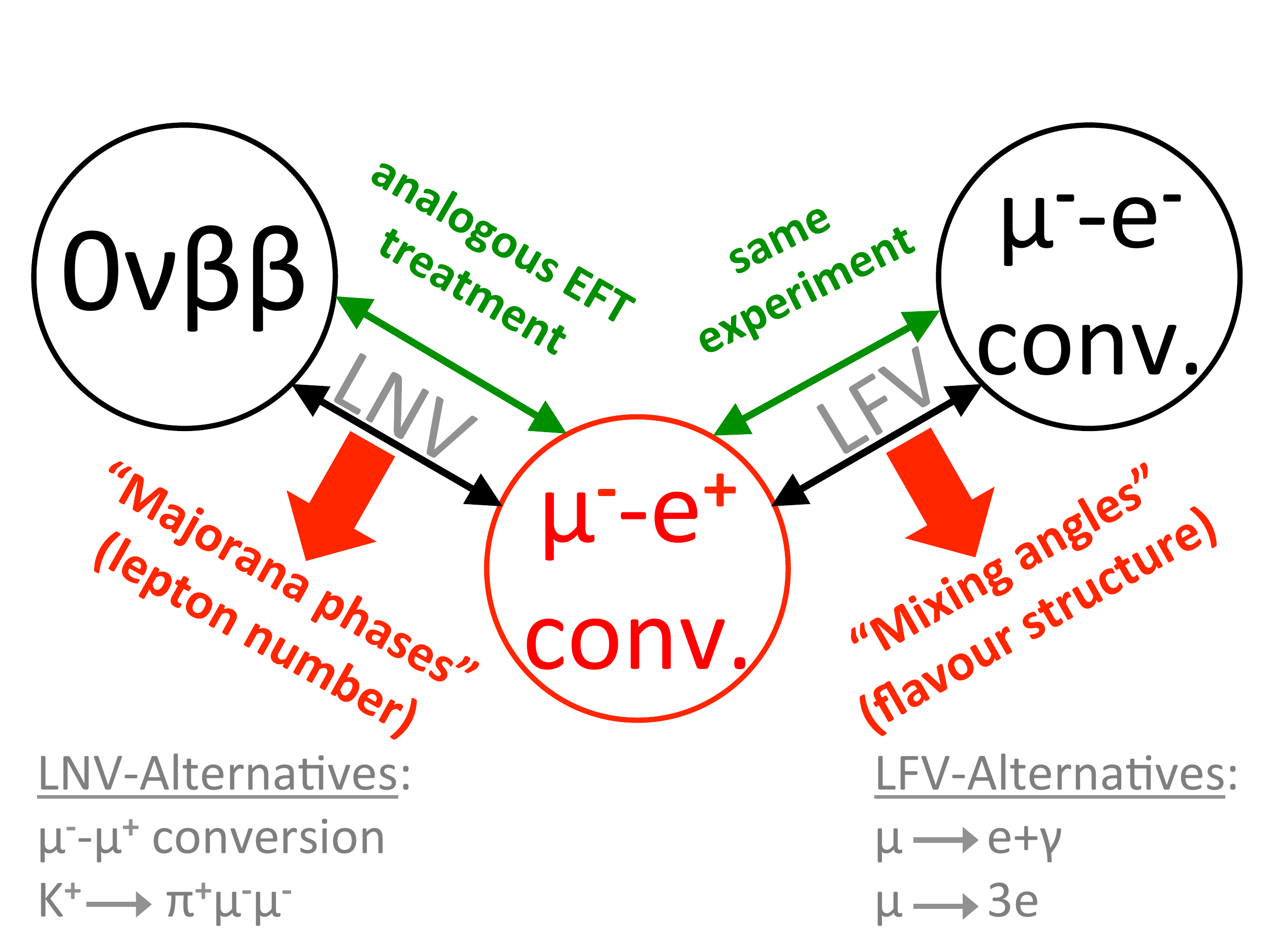}
  \caption{\label{fig:scheme}Schematic illustration of the profound role of $\mu^-$-- $e^+$ conversion.}
\end{figure}

In Fig.~\ref{fig:scheme}, we have illustrated how $\mu^-$-- $e^+$ conversion compares to both $\mu^-$-- $e^-$ conversion and $0\nu\beta\beta$: while $\mu^-$-- $e^-$ conversion can only detect lepton \emph{flavour} violation -- which we know exists from neutrino oscillation experiments -- and not the much more fundamental lepton \emph{number} violation, $0\nu\beta\beta$ can detect LNV but only in the $ee$-sector. Instead, $\mu^-$-- $e^+$ conversion is in some sense the best of both worlds, being able to detect LNV in the $e\mu$ sector. This is a great benefit given that there are models in which LNV is much more prevalent in flavour non-diagonal transitions. On top of that, and this is the actual candy, most experiments searching for ordinary $\mu^-$-- $e^-$ conversion can, even without modifications, \emph{at the same time} look for $\mu^-$-- $e^+$ conversion. This is next to be discussed.

In contrast to coherent $\mu^-$-- $e^-$ conversion, which can occur at a single nucleon and is mediated via the ground state of the nucleus, $\mu^-$-- $e^+$ conversion has to occur at two nucleons to allow for a $\Delta Q=2$ process (see section 3.5.1 of Ref.~\cite{Bernstein:2013hba} for a pedagogical summary of theoretical and experimental aspects). In this way it is very similar to $0\nu\beta\beta$, just with a muon instead of an electron. Taking the simplest case of Majorana neutrino exchange~\cite{Domin:2004tk}, the effective mass obtained from $0\nu\beta\beta$ contains terms proportional to $U_{ei}^2$, with $U_{ei}$ being the $i$-th element of the first row in the leptonic mixing matrix, whereas $\mu^-$-- $e^+$  conversion is sensitive to $U_{\mu i} U_{ei}$, therefore providing complementary information. Even if the process is not mediated by Majorana neutrino exchange, there will always be some connection between $\mu$ and $e$, which is absent for $0\nu\beta\beta$.

Experimentally, both types of bound muon conversion are two-step processes. First, a $\mu^-$ is captured in an atomic shell of higher principle quantum number, $n \approx 10$, before it quickly de-excites to the $1s$ ground state. The emission of the corresponding de-excitation photons (in case of muonic atoms this will be more than 100~keV in energy) serves as indicator for a shell capture. In case of Al (100\% of Al-27), which will be used in the next generation of experiments the $2p \ra 1s$ transition with the emission of a 346.8~keV photon with $79.7(6)\%$ intensity will serve as signal. After that the muon either decays in orbit (DIO), experiences a standard muon capture with the emission of a neutrino, or it undergoes $\mu$-- $e$ conversion in which it is captured by the nucleus and reemits a positron or electron. Assuming only "coherent"\footnote{Please note that the quotation marks are added here, since {\it coherent} technically refers to a process that has {\it the same} initial and final nucleus in the ground state, which cannot be the case for $\mu^-$-- $e^+$ conversion.} conversion -- which means that both initial and final state nucleus are in ground state -- the positron/electron created is fast, and it escapes the final-state atom. The positron/electron energy is then given by $E = m_\mu - B_\mu - E_{\rm rec}$, with $m_\mu$ being the muon mass, $B_\mu$ the binding energy of the $1s$-state in the muonic atom, and $E_{\rm rec}$ the nuclear recoil energy. The last two terms are small compared to the muon mass so that, in the exemplary case of Al-27, the expected energy of the electron is 104.97~MeV. 
\begin{figure}[t]
  \centering
  \includegraphics[trim=0cm 0cm 0cm 6cm,width=10cm]{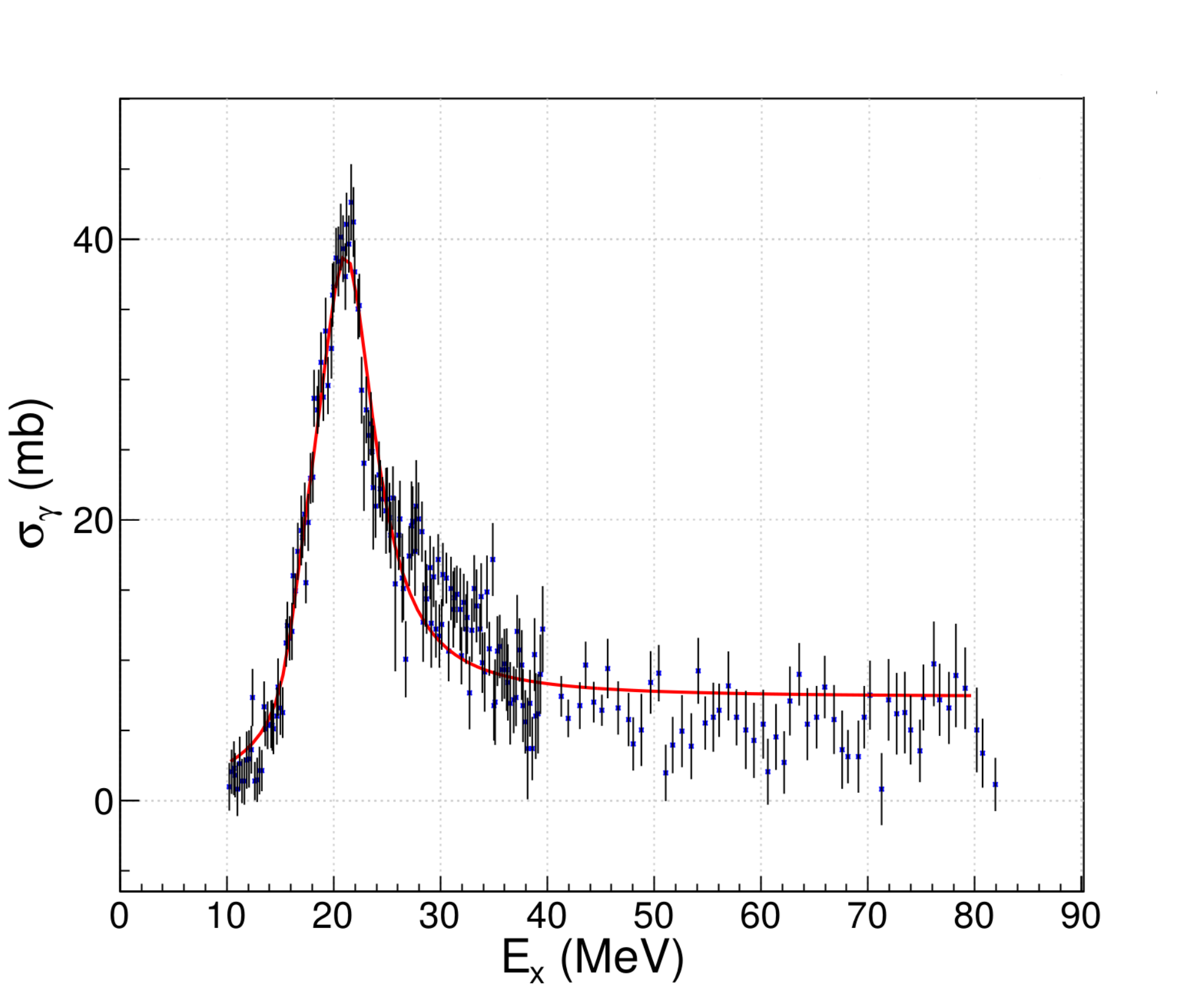}
  \caption{\label{fig:gdr}Shape of the giant dipole resonance in Al-27 using the EXFOR database. A Breit-Wigner shape is fitted to the data.}
\end{figure}

While $\mu^-$-- $e^-$ conversion is dominated by its coherent conversion~\cite{Kuno:1999jp}, this may be very different for the CLNFV $\mu^-$-- $e^+$ conversion, where several states can be excited and the resulting positrons will therefore have a more involved spectrum~\cite{Badertscher:1981ay}. Past measurements of $\mu^-$-- $e^+$ conversion~\cite{Bryman:1972rf,Abela:1980rs,Badertscher:1981ay,Burnham:1987gr,Ahmad:1988ur,Kaulard:1998rb}, the last one being SINDRUM~II~\cite{Bertl:2006up}, assumed that this process is completely mediated through the giant dipole resonance (GDR). SINDRUM~II used a Ti target and assumed a Breit-Wigner shape to fit the GDR with 20~MeV excitation energy and 20~MeV width. In the case of Al-27, which is the muon capture target for both future experiments COMET~\cite{COMET} and Mu2e~\cite{Kutschke:2011ux}, much better data exist and, using the EXFOR database~\cite{EXFOR}, the GDR can be fitted by a Breit-Wigner shape with a mean of 21.1~MeV and a width of 6.7~MeV, which is much more precise than the one used in the past (see Fig.~\ref{fig:gdr}). Hence, if this process is completely mediated by the GDR, which is an assumption, the positron energy will be 83.9 MeV and thus suffer from the higher background. The major background will be radiative muon and pion captures followed by asymmetric pair production, with the first process being dominant. The positron detection with respect to electrons should have a very high discrimination power due to the different orientation of the helical path in the magnetic
field.
 If the  $\mu^-$-- $e^+$ conversion is proceeding to a certain fraction via the ground state or via states between ground state and the GDR, then the signal will be smeared out over the range between the two values given. However, newer calculations revealed that a significant fraction (around to 40\%~\cite{Domin:2004tk}) are going via the ground state also for this process, which would be good news. Clearly this issue deserves future investigations to clarify how this process is mediated in a nucleus at all and whether the GDR is really playing a key role~\cite{Divari:2002sq,Domin:2004tk}.\\

\section*{Physics reach of COMET and similar experiments}

The goal of this section is to illustrate that experiments like COMET~\cite{COMET} could make a countable physics impact when aiming to measure $\mu^-$-- $e^+$ conversion. Note that, however, this section is only illustrative because currently several key pieces of information are not available yet: while for the known case of the operator $\epsilon^{xxz}_3$ upcoming experiments can by an inch \emph{not} scratch the surface of the relevant parameters, we will see that they are still close enough for investigations of the other possible operators to be interesting.

\begin{figure}[t]
 \centering
 \includegraphics[width=12cm]{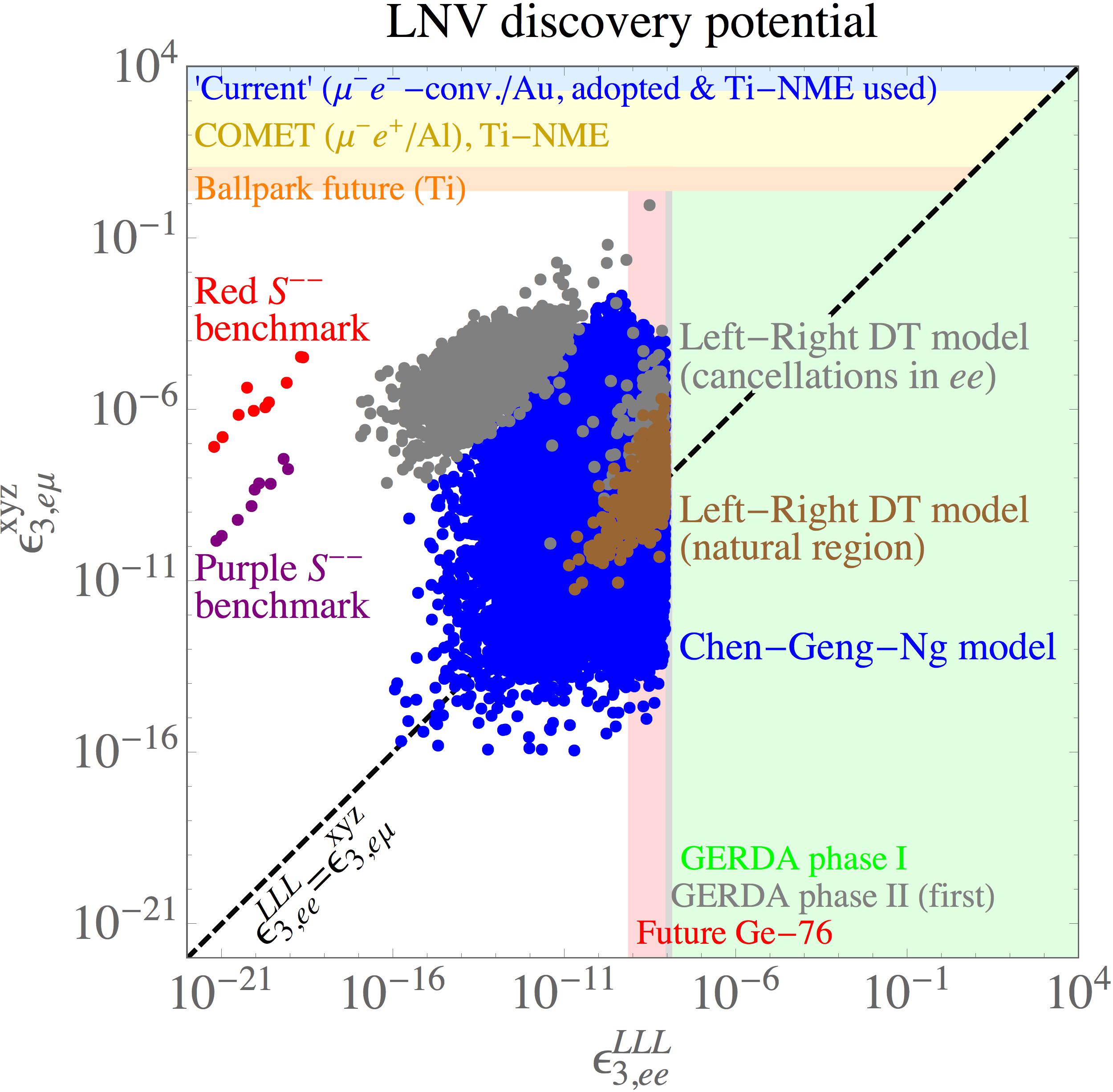}
 \caption{\label{fig:COMET-Reach}Illustration of the reach of future experiments for $\epsilon_3$.}
\end{figure}

To illustrate the potential of future experiments to detect LNV in the $e\mu$-sector, we display the limits on and sensitivities to the two effective parameters $\epsilon^{LLL}_{3,ee}$ and $\epsilon^{xyz}_{3,e\mu}$\footnote{Note that we allow for $x\neq y$ when illustrating the reach of future experiments, as argued for in Sec.~\ref{sec:formal}, although relying on the NMEs determined for $x=y$.} in Fig.~\ref{fig:COMET-Reach}, both for on-going and future experiments on $0\nu\beta\beta$ and on $\mu^-$-- $e^+$ conversion. In the former case, we illustrate the current limits for GERDA phase~I (light green region, from~\cite{Agostini:2013mzu}) and for the first data of phase~II (light grey slice, from~\cite{Matteo_Neutrino2016}), as well as a future projection of what could possibly be reached by experiments with Ge-76 (light red region, from~\cite{Agostini:2015dna}). In the case of $\mu^-$-- $e^+$ conversion, however, the information is much more scarce, as already mentioned. For example, up to now no experiment has used Al-27 to study ordinary $\mu^-$-- $e^-$ conversion, which is why there is no actual upper limit from that isotope. Other limits do exist, and for illustration we show how the bound on Au-197 from SINDRUM~II (light blue region, from~\cite{Bertl:2006up}; see~\cite{Geib:2015unm} for a collection of further limits) would translate into a limit on $\epsilon_{3,e\mu}$, provided that the sensitivity for $\mu^-$-- $e^-$ conversion is identical to that for $\mu^-$-- $e^+$ conversion (which is a good approximation up to a factor of $\mathcal{O}(1)$~\cite{Bryman:1972rf,Burnham:1987gr}). It is, however, important to keep in mind that the values of the NMEs are uncertain as stated before. While we would expect somewhat similar numbers for all isotopes, which is roughly the case for $0\nu\beta\beta$-NMEs~\cite{Menendez:2016kkg}, the only explicit value for $\mu^-$-- $e^+$ conversion mediated by heavy particles was $5.2$, computed in~\cite{Domin:2004tk} for Ti-48, as to be investigated by PRISM/PRIME (light orange region, from~\cite{Barlow:2011zza}). However, for Al-27 as used in COMET, we have not found a tabulated value, so that we had to rely on the value of $5.2$ at least serving as a ballpark estimate (light yellow region, from~\cite{COMET}). Clearly, the message is that further investigations from the nuclear physics side are needed.

In any case, the values used should serve as an illustration. What is clearly visible from the plot, though, is that the possible upper bounds from $0\nu\beta\beta$ on $\epsilon_{3,ee}$ are superior compared to those from $\mu^-$-- $e^+$ conversion on $\epsilon_{3,e\mu}$, by about eight to nine orders of magnitude at least. This is to be expected, since experiments on $0\nu\beta\beta$ can usually operate with a solid target while muon conversion experiments have to rely on high intensity muon beams, such that there is a massive enhancement of the former type of experiments by the Avogadro number. However, this is not the full picture, since there could be particle physics models in which much more LNV is contained in the $e\mu$- than in the $ee$-sector, i.e., their predictions would be situated in the upper left half of the plot. Although this information has not always been worked out, we have already in a first investigation been able to identify several models in which LNV is much more prominent in the $e\mu$-sector, depicted by the scattered points in Fig.~\ref{fig:COMET-Reach}, many of which are located in the upper left half of the plot. The examples displayed are the red and purple allowed benchmark points from a 2-loop neutrino mass model containing a doubly charged scalar~\cite{King:2014uha}, two regions from a Left-Right symmetric model supplemented by additional Higgs bosons and singlet fermions~\cite{Pritimita:2016fgr} (with the natural points depicted in brown and those which feature a cancellation in the $ee$-sector by the grey points), and an explicit example of $\mu^-$-- $e^+$ conversion mediated by a superposition of doubly charged singlet and triplet scalar components~\cite{Chen:2006vn} (blue points in the plot).\footnote{Note that we have already imposed the current bounds, which is why some sets of points seem to feature a sharp edge on the right.} While these models still cannot be probed by the upcoming conversion experiments, at least the grey points nearly peak into the region accessible by future experiments, thereby illustrating that valuable new information is likely to be reached for more suitable settings and/or other operators. In particular, some of the potential long-range contributions look promising~\cite{Babu:1995vh}.

Recall further that there are hardly any detailed investigations available at the moment, and we have only presented a few example models so that, in fact, there is potential to reveal further particle physics settings that can be probed by $\mu^-$-- $e^+$ conversion.

\section{\label{sec:conc}Summary and conclusions}

Having discussed several aspects of the process of $\mu^-$-- $e^+$ conversion, we can conclude that improvements are needed on three sides:
\begin{enumerate}

\item {\bf Experimentally}, no very detailed sensitivity studies exist at this stage for $\mu^-$-- $e^+$ conversion. Although new backgrounds such as protons~\cite{Abela:1980rs} and pions~\cite{Badertscher:1981ay} may apear, we can nevertheless expect sensitivity levels to be at least similar to those on $\mu^-$-- $e^-$ conversion~\cite{Bryman:1972rf,Ahmad:1988ur,Badertscher:1981ay}, with some previous limits being nearly identical for both processes~\cite{Bryman:1972rf,Burnham:1987gr}. Such studies can and should be done with existing resources.

\item {\bf Nuclear matrix elements} have hardly been computed for $\mu^-$-- $e^+$ conversion, which is particularly true for Al-27. The only two available are those for $\epsilon^{xxz}_3$ with light ($\mathcal{M}_\nu = 0.025$) and heavy ($\mathcal{M}_N = 5.2$) neutrino mediation~\cite{Simkovic:2000ma,Domin:2004tk}, both for Ti-48, with the latter being equivalent to the NME needed for the short-range $\epsilon_3$-operator from Eq.~\eqref{eq:short-range}. However, no further up-to-date computations seem to exist, \emph{neither} for isotopes other than Ti-48 \emph{nor} for short-range operators other than $\epsilon^{xxz}_3$.\footnote{There is in fact one computation of the NME available for the case of Al-27~\cite{Divari:2002sq}, however, that references only treated the possibly unobservable~\cite{Merle:2006du} case of light neutrino mediation, but no short-range operators -- and at least for Ti-48, this result did not carry over to heavy mediators~\cite{Simkovic:2000ma,Domin:2004tk}.} Furthermore, there seem to exist no theoretical works investigating which percentage of $\mu^-$-- $e^+$ conversion takes place via the  "ground state $\to$ ground state" transition and how this influences the discovery potential. This makes it currently \emph{impossible} to present the full picture. Indeed, glancing at Fig.~\ref{fig:COMET-Reach}, it seems realistic that some models may be promising, and further investigations could reveal settings in reach of experiments. In~\cite{Geib:2016daa}, we will identify several other contributions realising, e.g., operators $\epsilon_1$ and $\epsilon_2$. These constributions do look rather promising, and they may in fact have a greater potential to be detected in the near future. However, without any computations of the NMEs, this cannot be judged. We would therefore like to transmit this message to the nuclear theory community since, in fact, LNV could possibly be found in $\mu^-$-- $e^+$ conversion more easily than in $0\nu\beta\beta$. \emph{Getting a better understanding of the nuclear physics part is the most important ingredient to make progress on $\mu^-$-- $e^+$ conversion.}

\item On the {\bf particle physics} side, there are for many models no detailed studies on how much LNV could be present in the $e\mu$-sector. There exist detailed studies on $0\nu\beta\beta$ though, see e.g.\ Refs.~\cite{Deppisch:2015cla,Simkovic:2010ka}, which has been the focus for years. But, as we have illustrated, LNV in the $ee$-sector may be suppressed. However, most cases are only studied superficially, while new options like $\mu^-$-- $e^+$ conversion can be available but are not discussed in detail. A comprehensive study on $\mu^-$-- $e^+$ conversion from a technical point of view including the derivation of the decay rate in Eq.~(\ref{eq:DecayRate}) and the investigation of several LNV models is currently in preparation~\cite{Geib:2016daa}.

\end{enumerate}

Summing up, we are in a position in which experiments have a great potential to advance our knowledge on CLNFV in the $e\mu$-sector. However, the theory side has to gain momentum, both for particle and nuclear physics, since beneficial steps are obvious but not made. Only if all three communities pull together, advances will be achieved.

\section*{Acknowledgements}

We would like to thank D.~Dercks, R.~Schwengner, F.~Simkovic, and T.~Wester for useful discussions. AM acknowledges partial support by the the Micron Technology Foundation, Inc., as well as by the European Union's Horizon 2020 research and innovation programme under the Marie Sklodowska-Curie grant agreements No.~690575 (InvisiblesPlus RISE) and No.~674896 (Elusives ITN).



\bibliographystyle{./apsrev}
\bibliography{MuCon}

\end{document}